\documentclass{pos}

\usepackage{amsmath,amssymb}

\newcommand{\be}{\begin{equation}}
\newcommand{\en}{\end{equation}}
\newcommand{\bi}{\begin{itemize}}
\newcommand{\ei}{\end{itemize}}
\newcommand{\bea}{\begin{eqnarray}}
\newcommand{\ena}{\end{eqnarray}}

\newcommand{\hbo}{\hbox to 1 true cm {\hfill } }
\newcommand{\tr}{\hbox{tr}}

\title{Solution of the Gribov problem from gauge invariance }

\ShortTitle{Solution of the Gribov problem from gauge invariance }

\author{\speaker{Kurt Langfeld} \\
        School of Mathematics \& Statistics, University of Plymouth,
        Plymouth, PL4 8AA, UK  \\
        E-mail: \email{kurt.langfeld@plymouth.ac.uk}}

\author{Tom Heinzl \\
        School of Mathematics \& Statistics, University of Plymouth,
        Plymouth, PL4 8AA, UK }
\author{Anton Ilderton \\
        School of Mathematics \& Statistics, University of Plymouth,
        Plymouth, PL4 8AA, UK {\it and } \\
        School of Mathematics, Trinity College, Dublin 2,
        Ireland }
\author{Martin Lavelle \\
        School of Mathematics \& Statistics, University of Plymouth,
        Plymouth, PL4 8AA, UK }
\author{David McMullan \\
        School of Mathematics \& Statistics, University of Plymouth,
        Plymouth, PL4 8AA, UK }

\abstract{A new approach to gauge fixed Yang-Mills theory is derived using the Polyakov-Susskind projection techniques to build gauge invariant
states. In our approach, in contrast to the Faddeev-Popov method, the Gribov problem does not prevent the gauge group from being factored out of the partition function. Lattice gauge theory is used to
illustrate the method via a calculation of the static quark--antiquark
potential generated by the gauge fields in the fundamental modular
region of Coulomb gauge.
}

\FullConference{8th Conference Quark Confinement and the Hadron Spectrum\\
                 September 1-6, 2008\\
                 Mainz. Germany}

\begin{document}

\section{Introduction}
Yang-Mills theories are the cornerstone of the standard model.
These theories are formulated in terms of a highly redundant set
of variables, due to the {\it local} gauge symmetry of the action.
In case of Yang-Mills theories with matter, the variables are the
gauge fields $A_\mu(x)$ and the matter fields $Q(x)$.
The calculation of physical observables is hampered by the fact that
these fields are not directly related to physical particles.
This can be easily seen from the fact that they transform non-trivially under
gauge transformations $\Omega$ in the gauge group $\cal G$:
\be
	A_\mu(x) \rightarrow A_\mu ^\Omega (x) , \hbo
Q(x)  \rightarrow \Omega(x) \, Q(x) .
\label{eq:1}
\en
Gauge invariance tells us that we can work with any representative of the gauge orbit $\{ A_\mu^\Omega (x) \vert \Omega (x) \; \in {\cal G} \}$. A common method of choosing a {\it unique } representative is through the {\bf global} maximum of a given gauge fixing action:
\be
S_\mathrm{fix} [A^\Omega ] \rightarrow \hbox{global maximum}
\hbo \hbox{for} \hbo \Omega \; =: \;  \Omega _\mathrm{FMR}[A].
\label{eq:2}
\en
The set of these representatives is free of Gribov copies and
(after suitable boundary identifications)
is called the {\it fundamental modular region}:
\be
\left\{ A_\mu^\Omega (x) \, \vert \, \Omega  = \Omega _\mathrm{FMR}[A]
\right\}.
\label{eq:3}
\en
It has been known for a long time~\cite{Dirac:1955uv}
that a physical electron state
$\vert Q \rangle $ can only be obtained if the bare state $Q(x) \vert 0
\rangle $ is properly dressed by a photon cloud:
\be
\vert Q \rangle \, = \, h[A] \, Q(x) \vert 0 \rangle \, , \hbo
h[A^\Omega ] \, = \, h[A] \, \Omega ^\dagger \; .
\label{eq:4}
\en
The latter equation specifies the dressing property which
ensures the gauge invariance of $\vert Q \rangle $. The choice of a
a dressing is not unique. They may differ e.g.\ 
by a gauge invariant factor without spoiling the dressing property
in (\ref{eq:4}). One particular way to construct a dressing is to use
gauge fixing:
\be
h[A] \, = \, \Omega _\mathrm{FMR}[A]  \; .
\label{eq:5}
\en
We call this definition {\it dressing from the fundamental modular region}.
Under the assumption that there is a unique global maximum,
it can be easily shown that this definition satisfies the dressing
property:
$$
h[A^\Omega ] \, = \, \Omega _\mathrm{FMR}[A^\Omega ] =
\Omega _\mathrm{FMR}[A]  \, \Omega ^\dagger \; = \; h[A] \, \Omega ^\dagger \; .
$$
The disadvantage of this definition is that it relies on an explicit
realisation of gauge fixing. While the Faddeev-Popov method~\cite{Faddeev:1967fc} works in perturbation theory, it fails non-perturbatively, leading to Green functions being in indeterminate
form~\cite{Baulieu:1996kb,Baulieu:1996rp,Neuberger:1986xz}. This is
because in non-Abelian theories the gauge fixing condition
has multiple solutions $\Omega (x)$.
This was first elucidated by Gribov for the case of Coulomb
gauge~\cite{Gribov:1977wm}, and since then has been known as the {\it Gribov problem}.
One might argue that in a purely  numerical approach using lattice
regularisation there is no need to resort to a gauge fixing condition:
 one might find the unique representative on the gauge orbit by
seeking the global maximum of the gauge fixing action (\ref{eq:2})
using advanced algorithms. While this approach is conceptually correct,
it is not feasible: it can be shown that locating the global
maximum is equivalent to solving a spin-glass problem which is beyond the
scope of the numerical techniques currently available.

\vskip 0.3cm
In order to bypass the Gribov problem, we recently
proposed~\cite{Heinzl:2008bu} combining the
alternative construction of the gauge invariant partition function by
Parrinello and
Zwanziger~\cite{Parrinello:1990pm,Zwanziger:1990tn,Dudal:2008sp}  with the technique of gauge invariant projection initiated by Polyakov and
Susskind~\cite{Polyakov:1978vu,Susskind:1979up,Gross:1980br}.
This leads to
our {\it integral dressing} construction of
gauge invariant trial states, which we will describe below. We will also describe the ``ice-limit'', in which we recover
the dressing from the FMR {\it without having to construct the FMR
explicitly}. As an initial application, we employ the integral dressed
trial states to
calculate the static interquark potential.

\section{Integral dressing }

In the following, we will adopt lattice regularisation. The gauge
fields $A_\mu(x)$ are represented by group valued links
$U_\mu (x)$.
Rather than realising the dressing property (\ref{eq:4}) by gauge fixing,
we define the integral dressing~\cite{Heinzl:2008bu}
\be
h[U] = \int {\cal D} \Omega \; \; \Omega (x) \;
\exp \{ \kappa \;  S_\mathrm{fix}[U ^\Omega ] \} ,
\label{eq:6}
\en
where $\kappa $ plays the role of a gauge fixing parameter.
Using the gauge invariance of the Haar measure, we easily verify the
dressing property $(G\in {\cal G})$:
\bea
h[U^{G} ] &=& \int {\cal D} \Omega \; \; \Omega (x) \;
\exp \{ \kappa \;  S_\mathrm{fix}[U ^{\Omega  G}] \}
\; = \; \int {\cal D} [ \Omega { G} ] \; \; [ \Omega { G}]  \;
{ G^\dagger } \;
\exp \{ \kappa \;  S_\mathrm{fix}[U ^{\Omega  G}] \}
\; = \; h[U] \, { G^\dagger } .
\nonumber
\ena
Gauge invariant dressings for multi-quark trial states may be defined similarly, e.g. for a quark--antiquark trial state we make the ansatz
\be
 \vert \mathrm{trial} \rangle :=
Q^\dagger (y) \; h^{(2)}[U](y,x) \; Q(x) \; \vert 0 \rangle , \hbo
h^{(2)}[U] :=
\int {\cal D} \Omega \; \; \Omega ^\dagger (y) \; \Omega
(x) \; \exp \{ {\kappa} \;  S_\mathrm{fix}[U ^\Omega ] \} .
\label{eq:7}
\en
Let us briefly discuss the {\it strong gauge fixing limit} $\kappa
\to \infty $. In this case, the total contribution to the integral
in (\ref{eq:7}) arises solely from the domain where
$S_\mathrm{fix}$ attains its global maximum. In this case, we recover
the dressing from the FMR:
\be
h^{(2)}[U] \; \propto \; \Omega _{_\mathrm{FMR}}^\dagger (y) \;
\Omega _{_\mathrm{FMR}} (x) , \hbo
\vert FMR \rangle \, := \, \lim _{\kappa \to \infty } \vert \mathrm{trial}
\rangle .
\label{eq:8}
\en
As an illustrative example, we consider {\bf axial dressing} by choosing
\begin{equation}\label{ineq}
 S_\mathrm{fix}[U ^\Omega ] \; = \; \sum _x \frac{1}{N_c}\,
\tr \, U^\Omega_3(x) \; \le \;
N_x \; ,
\end{equation}
where $N_x$ is the number of lattice cites, and $N_c$ is the number of
colours of the gauge group. The key observation is that the upper bound in (\ref{ineq}) can be saturated exactly: letting ${\cal C}_x$ denote
the straight line joining the points $(x_1,x_2,1,x_4)$ and
$(x_1,x_2,x_3,x_4)$, we find
$
\Omega _\mathrm{FMR} (x) \; = \; \prod _{{\cal C}_x} U_3(x) 
$
in the strong gauge fixing limit of $\kappa\to\infty$.
The integral dressing (\ref{eq:8}) for a quark at $x$ and an antiquark
at $y=x + r \, e_3$ then becomes
\be
h^{(2)}[U](y,x) \; = \; \prod _{C_{xy}} U_3 (x) ,
\label{eq:9}
\en
where $C_{xy}$ is the straight line between $x$ and $y$.
Relation (\ref{eq:9}) and the corresponding trial state have an intuitive
interpretation: $h^{(2)}$ is just the Polyakov line between the points
$x$ and $y$, so the trial state is made gauge invariant by joining the quark and antiquark with a thin gluonic string.

\bigskip
The static quark antiquark potential $V(r)$ can be obtained by
calculating the expectation value of the (Euclidean) time
evolution operator using the FMR trial state:
\be
\langle \mathrm{trial} \vert \mathrm{e}^{- H T } \vert \mathrm{trial} \rangle
\stackrel{T \, \mathrm{large}}{\longrightarrow }
\vert \langle 2 \vert  \mathrm{trial} \rangle \vert ^2
\; \mathrm{e}^{ - V(r) T }, \hbo
\langle \mathrm{FMR} \vert \mathrm{e}^{- H T } \vert \mathrm{FMR} \rangle
= W(r,T) .
 \label{eq:10}
\en
In the case of the axial dressing, we recover the standard approach
to the static potential using rectangular Wilson loops $W(r,t)$.
Although we have not encountered here a Gribov problem in the case of axial dressing,
its disadvantage is that the overlap $\vert \langle 2 \vert
\mathrm{trial} \rangle \vert ^2 $ is poor and even vanishes when the
lattice regulator is removed~\cite{Heinzl:2008tv,Heinzl:2007kx}.

\section{Dressing from the Coulomb gauge FMR - the ice-limit }
\begin{figure}
\centerline{
\includegraphics[width=.4\textwidth]{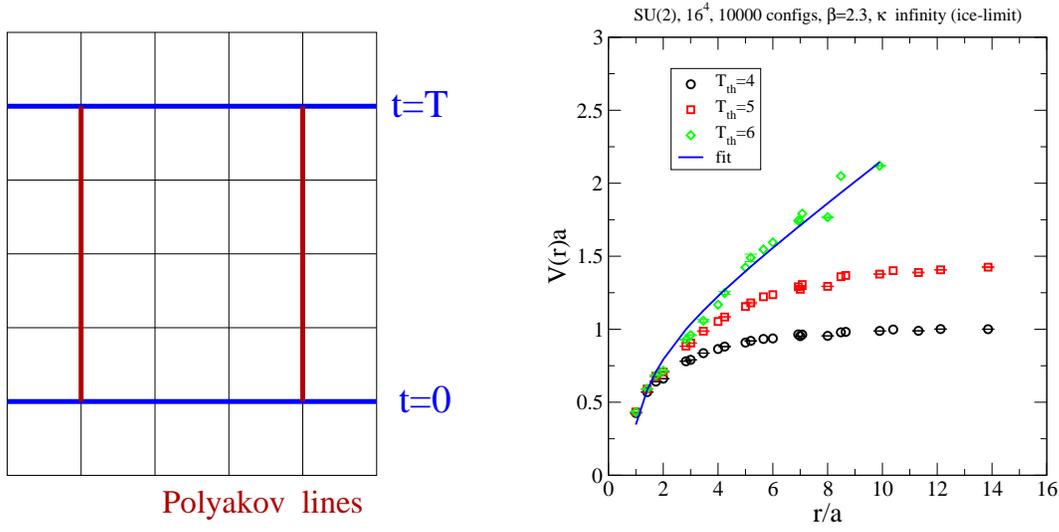} \hspace{1cm}
\includegraphics[width=.45\textwidth]{pot23_2.eps}
}
\caption{Illustration of the ice-limit (left panel) and
the static potential from Coulomb dress trial states (right panel).
}
\label{fig1}
\end{figure}
A much better overlap is observed if Coulomb gauge fixing
is employed~\cite{Heinzl:2008tv}:
\begin{equation}\label{coul}
 S_\mathrm{fix}[U ^\Omega ] \; = \; \sum _{k,x} \frac{1}{N_c}
\, \tr \, U^\Omega_k(x) \; , \hbo k = 1 \ldots 3 .
\end{equation}
Using the definition of the integral dressing and heavy quark
propagation, the expectation value (\ref{eq:10})
can be written as~\cite{Heinzl:2008bu}:
\be
\langle \mathrm{trial} \vert \mathrm{e}^{- H T } \vert \mathrm{trial}
\rangle \propto \int {\cal D} U_\mu \; {\cal D} \Omega \; \ldots \;
\exp\{ { \kappa S_\mathrm{fix}[U^\Omega]} \} \; .
\label{eq:11}
\en
Note the order of the integration: the $\Omega $ integration is done before
the link $U_\mu $ integration. With this ordering, we recover the Gribov problem in the
large-$\kappa $ limit, since for a given background field $U_\mu (x)$,
the $\Omega $-path integral may be viewed as a spin-glass partition function
with a metric $U_\mu (x)$. However, using the explicit gauge invariance
of our approach, the order of integration can be
exchanged, and the $U$ integral becomes independent of $\Omega$ (for details see ~\cite{Heinzl:2008bu}),
\be
\langle \mathrm{trial} \vert \mathrm{e}^{- H T } \vert \mathrm{trial}
\rangle \propto \left( \int {\cal D} \Omega \right) \;
\int {\cal D} U_\mu \; \ldots \;
\exp\{ \kappa S_\mathrm{fix}[U] \} .
\label{eq:20}
\en
Changing the order of integration solves the Gribov problem:
the limit $\kappa \to \infty $ can thus be taken analytically, as the dominant contributions to the integral now arise from maximising (\ref{coul}) with respect to $U$ rather than $\Omega$. The solution is clearly to take $U=1$, i.e.
\be
 S_\mathrm{fix} [U] \stackrel{U_\mu }{\rightarrow }
\hbox{max}\hbo \implies\hbo
U_k (\vec{x},t) =1 \; \; \hbox{for} \; \; t=0
\; \; \hbox{and} \; \; t=T \; .
\label{eq:21}
\en
The limit $\kappa \to \infty $ implies that the link fields $U_\mu $ on the time-slices
$t=0$ and $t=T$ are frozen to perturbative vacuum levels.
We have therefore called this the {\it ice-limit}.
An illustration of the ice-limit together with our numerical results for
$V(r)$ can be found in figure~\ref{fig1} -- a linearly rising
interquark potential clearly emerges.

\end{document}